\documentstyle[preprint,aps,prbbib,graphicx]{revtex}
\newcounter{newpart}
\makeatletter
\newcommand{\mynewpart}{\arabic{newpart}}
  \renewcommand{\thefigure}{\mynewpart.\@arabic\c@figure}
  \makeatother
  
\begin{document}
\thispagestyle{empty}
\begin{center}{COSMOLOGICAL RELATIVITY: 
A NEW THEORY OF COSMOLOGY}\end{center}
\vspace{1cm}
\begin{center}{Silvia Behar and Moshe Carmeli}\end{center} 
\begin{center}{Department of Physics, Ben Gurion University, Beer Sheva 84105, 
Israel}\end{center}
\begin{center}{(E-mail: carmelim@bgumail.bgu.ac.il \hspace{1cm} 
silviab@bgumail.bgu.ac.il)}\end{center}
\begin{center}{ABSTRACT}\end{center}
A general-relativistic theory of cosmology, the dynamical variables of 
which are those of Hubble's, namely distances and redshifts, is presented.
The theory describes the universe as having a three-phase evolution with a
decelerating expansion followed by a constant and an accelerating expansion,
and it predicts that the universe is now in the latter phase. The theory is
actually a generalization of Hubble's law taking gravity into account by means 
of Einstein's theory of general relativity. The equations obtained for the 
universe expansion are elegant and very simple. It is shown, assuming 
$\Omega_0=0.24$, that the time at which the universe goes over from a 
decelerating to an accelerating expansion, i.e. the constant expansion phase,
occurs at 0.03$\tau$ from the big bang, where $\tau$ is the Hubble time in
vacuum. Also, at that time the cosmic radiation temperature was 11K. Recent 
observations of distant supernovae imply, in defiance of expectations, that
the universe growth is accelerating, contrary to what has always been assumed 
that the expansion is slowing down due to gravity.Our theory confirms these
recent experimental results by showing that the universe now is definitely in 
a stage of accelerating expansion.
\newpage
\section{Introduction}
In this paper we present a theory of cosmology that is based on Einstein's
general relativity theory. The theory is formulated in terms of directly 
measured quantities, i.e. distances, redshifts and the matter density of the
universe.

The general-relativistic theory of cosmology started in 1922 with the 
remarkable work of A. Friedmann [1,2], who solved the Einstein gravitational 
field equations and found that they admit non-static cosmological solutions
presenting an expanding universe. Einstein, believing that the universe should 
be static and unchanged forever, suggested a modification to his gravitational 
field equations by adding to them the so-called cosmological term which can
stop the expansion.

Soon after that E. Hubble [3,4] found experimentally that the far-away galaxies 
are receding from us, and that the farther the galaxy the bigger its velocity.
In simple words, the universe is indeed expanding according to a simple
physical law that gives the relationship between the receding velocity and the 
distance,
$$\mbox{\bf v}=H_0\mbox{\bf R}.\eqno(1.1)$$
Equation (1.1) is usually referred to as the Hubble law, and $H_0$ is called
the Huble constant. It is tacitly assumed that the velocity is proportional
to the actual measurement of the redshift $z$ of the receding objects by
using the non-relativistic relation $z=v/c$, where $c$ is the speed of light 
in vacuum. 

The Hubble law does not resemble standard dynamical physical laws that are
familiar in physics. Rather, it is a {\it cosmological equation of state} of
the kind one has in thermodynamics that relates the pressure, volume and 
temperature, $pV=RT$ [5]. It is this Hubble's equation of state that will be
extended so as to include gravity by use of the full Einstein theory of
general relativity. The obtained results will be very simple, expressing 
distances in terms of redshifts; depending on the value of $\Omega=
\rho/\rho_c$ we will have accelerating, constant and decelerating expansions,
corresponding to $\Omega<1$, $\Omega=1$ and $\Omega>1$, respectively. But the
last two cases will be shown to be excluded on physical evidence, although the
universe had decelerating and constant expansions before it reached its
present accelerating expansion stage. As is well known the standard FRW 
cosmological theory does not deal directly with Hubble's measured quantities, 
the distances and redshifts. Accordingly, the present theory can be compared 
directly with important recent observations made by astronomers which defy
expectations.

In Sections 2 and 3 we review the standard Friedmann and Lema\^\i tre models. In 
Section 4 we mention some points in those theories. In Section 5 we 
present our cosmological theory written in terms of distances and redshifts,
whereas Section 6 is devoted to the concluding remarks.
\section{Review of the Friedmann Models}
Before presenting our theory, and in order to fix the notation, we very 
briefly review the existing theory [6,7]. In the four-dimensional curved
space-time describing the universe, our spatial three-dimensional space is
assumed to be isotropic and homogeneous. Co-moving coordinates, in which 
$g_{00}=1$ and $g_{0k}=0$, are employed [8,9]. Here, and throughout our paper,
low-case Latin indices take the values 1, 2, 3 whereas Greek indices will take
the values 0, 1, 2, 3, and the signature will be $(+---)$. The four-dimensional
space-time is split into $1\bigoplus 3$ parts, and the line-element is 
subsequently written as
$$ds^2=dt^2-dl^2,\hspace{1cm} dl^2={}^{(3)}
g_{kl}dx^k dx^l=-g_{kl}dx^k dx^l,\eqno(2.1)$$ and the $3\times 3$ tensor ${}^{(3)}g_{kl}
\equiv -g_{kl}$ describes the geometry of the three-dimensional space at a 
given instant of time. In the above equations the speed of light $c$ was taken 
as unity.

Because of the isotropy and homogeneity of the three-geometry, it follows that
the curvature tensor must have the form
$${}^{(3)}R_{mnsk}=K\left[{}^{(3)}g_{ms}{}^{(3)}g_{nk}-{}^{(3)}g_{mk}
{}^{(3)}g_{ns}\right],\eqno(2.2)$$   
where $K$ is a constant, the curvature of the three-dimensional space, which 
is related to the Ricci scalar by ${}^{(3)}R=-6K$ [10]. By simple geometrical
arguments one then finds that
$$dl^2=\left(1-r^2/R^2\right)^{-1}dr^2+r^2\left(d\theta^2+\sin^2\theta d\phi^2
\right),\eqno(2.3)$$
where $r<R$. Furthermore, the curvature tensor corresponding to the metric (2.3)
satisfies Eq. (2.2) with $K=1/R^2$. In the ``spherical" coordinates $\left(t,r,
\theta,\phi\right)$ we thus have
$$g_{11}=-\left(1-r^2/R^2\right)^{-1}.\eqno(2.4)$$ 
$R$ is called the radius of the curvature (or the expansion 
parameter) and its value is determined by the Einstein field equations.

One then has three cases: (1) a universe with positive curvature for which
$K=1/R^2$; (2) a universe with negative curvature, $K=-1/R^2$; and (3) a 
universe with zero curvature, $K=0$. The $g_{11}$ component for the 
negative-curvature universe is given by
$$g_{11}=-\left(1+r^2/R^2\right)^{-1},\eqno(2.5)$$
where $r<R$. For the zero-curvature universe one lets $R\rightarrow\infty$.

In this theory one has to change variables in order to get 
the
solutions of the Einstein field equations according to the type of the 
universe. Accordingly, one makes the substitution $r=R\sin\chi$ for the
positive-curvature universe, and $r=R\sinh\chi$ for the negative-curvature
universe. The time-like coordinate is also changed into 
another one $\eta$ by the transformation $dt=Rd\eta$. The corresponding line
elements then become:
$$ds^2=R^2\left(\eta\right)\left[d\eta^2-d\chi^2-\sin^2\chi\left(d\theta^2
+\sin^2\theta d\phi^2\right)\right]\eqno(2.6a)$$
for the positive-curvature universe,
$$ds^2=R^2\left(\eta\right)\left[d\eta^2-d\chi^2-\sinh^2\chi\left(d\theta^2
+\sinh^2\theta d\phi^2\right)\right]\eqno(2.6b)$$
for the negative-curvature universe, and
$$ds^2=R^2\left(\eta\right)\left[d\eta^2-dr^2-r^2\left(d\theta^2
+\sin^2\theta d\phi^2\right)\right]\eqno(2.6c)$$
for the flat three-dimensional universe. In the sequel, we will see that the
time-like coordinate in our theory will take one more different form. 

The Einstein field equations are then employed in order to determine the
expansion parameter $R(\eta)$. In fact only one field equation is needed,
$$R_0^0-\frac{1}{2}\delta_0^0 R+\Lambda\delta_0^0=8\pi GT_0^0,\eqno(2.7)$$
where $\Lambda$ is the cosmological constant, and $c$ was taken as unity. In
the Friedmann models one takes $\Lambda=0$, and in the comoving coordinates 
used one easily finds that $T_0^0=\rho$, the mass density. While this choice of
the energy-momentum tensor is acceptable in standard general relativity and in
Newtonian gravity, we will argue in the sequel that it is not so for 
cosmology. At any rate, using $\rho\left(t\right)=M/2\pi^2R^3$, where $M$ is 
the ``mass" and $2\pi^2R^3$ is the ``volume" of the universe, one obtains
$$3\left[\left(dR/dt\right)^2+1\right]/R^2=4GM/\pi R^3+\Lambda,\eqno(2.8a)$$
or, in terms of $\eta$ along with taking $\Lambda=0$,
$$3\left[\left(dR/d\eta\right)^2+R^2\right]/R=4GM/\pi.\eqno(2.9a)$$ 
The solution of this equation is
$$R={}^\star R\left(1-\cos\eta\right),\eqno(2.10a)$$
where ${}^\star R=2GM/3\pi$, and from $dt=Rd\eta$ we obtain
$$t={}^\star R\left(\eta-\sin\eta\right).\eqno(2.11a)$$
Equations (10a) and (11a) are those of a cycloid, and give a full 
representation for the expansion parameter of the universe. Fig. 2.1 shows a 
plot of $R$ as a function of $t$. At $t=0$, $\pm 2\pi{}^\star R$, 
$\pm 4\pi{}^\star R$, $\dots$, etc., $R(t)$ vanishes; that is, the universe
contracts to a point. Since the density will become very large when this is
about to happen, our approximate expression for the energy-momentum tensor 
will fail. We should also keep in mind that the classical Einstein equation
becomes inapplicable at very high densities. It is therefore not clear exactly
what happens at the singular points of Fig. 2.1, and we do not know whether
the universe actually has the periodic behavior suggested by this figure. 

Similarly, one obtains for the negative-curvature universe the analog to Eqs.
(2.8a) and (2.9a),
$$3\left[\left(dR/dt\right)^2-1\right]/R^2=4GM/\pi R^3+\Lambda,\eqno(2.8b)$$
$$3\left[\left(dR/d\eta\right)^2-R^2\right]/R=4GM/\pi,\eqno(2.9b)$$ 
the solution of which is given by 
$$R={}^\star R\left(\cosh\eta-\eta\right),\eqno(2.10b)$$
$$t={}^\star R\left(\sinh\eta-\eta\right).\eqno(2.11b)$$
Fig. 2.2 shows $R$ as a function of $t$. The universe begins with a big bang
and continues to expand forever. As $t\rightarrow\infty$, the universe 
gradually becomes flat. Again, the state near the singularity at $t=0$ is not
adequately described by our equations.

Finally, for the universe with a flat three-dimensional space the Einstein 
field equations yield the analog to Eqs. (2.8a) and (2.9a),
$$3\left(dR/dt\right)^2/R^2=4GM/\pi R^3+\Lambda,\eqno(2.8c)$$
$$3\left(dR/d\eta\right)^2/R=4GM/\pi.\eqno(2.9c)$$ 
As a function of $t$, the solution is
$$R=\left(3GM/\pi\right)^{1/3}t^{2/3}.\eqno(2.10c)$$  
This function is plotted in Fig. 2.3. As $t\rightarrow\infty$, the 
four-geometry tends to become flat.
\section{Lema\^\i tre Models} 
An extension of the Friedmann models was carried out by Lema\^\i tre, who
considered universes with zero energy-momentum but with a non-zero 
cosmological constant. While these models are of interest mathematically
they have little, if any, relation to the physical universe because we know
that there is baryonic matter. The behavior of the universe in this model will 
be determined by the cosmological term; this term has the same effect as a
uniform mass density $\rho_{eff}=-\Lambda/4\pi G$, which is constant in space
and time. A positive value of $\Lambda$ corresponds to a negative effective
mass density (repulsion), and a negative value of $\Lambda$ corresponds to a
positive mass density (attraction). Hence, we expect that in the universe with
a positive value of $\Lambda$, the expansion will tend to accelerate; whereas
in a universe with negative value of $\Lambda$, the expansion will slow down,
stop, and reverse.

The equations of motion for $R(t)$ have been derived in Section 2, but here
it will be assumed that $\Lambda\neq 0$ whereas the energy-momentum tensor 
appearing in Eq. (2.7) to be zero. For the positive-curvature universe one 
obtains the analog to Eq. (2.8a),
$$3\left[\left(dR/dt\right)^2+1\right]/R^2=\Lambda.\eqno(3.1a)$$
From equation (3.1a) one immediately concludes that $-1+\Lambda R^2/3$ cannot
be negative. This implies that $\Lambda>0$, and that the value of $R$ can
never be less than $(3/\Lambda)^{1/2}$, i.e. the radius of curvature cannot
be zero which excludes the possibility of a big bang.

The integration of Eq. (3.1a) yields,
$$R\left(t\right)=\left(3/\Lambda\right)^{1/2}\cosh\left[\left(\Lambda/3
\right)^{1/2}t\right] \eqno(3.2a)$$
where $t$ was taken zero when $R$ has its minimum value. Fig. 3.1, curve (a), 
shows a plot of $R$ as a function of $t$. As is seen, for $t>0$ the universe
expands monotonically, and as $t$ increases, $R$ increases too and the 
universe becomes flat.

Similarly, one obtains for the negative-curvature universe the analog to Eq.
(3.1a),
$$3\left[\left(dR/dt\right)^2-1\right]/R^2=\Lambda,\eqno(3.1b)$$
the integration of which gives,
$$R\left(t\right)=\left(3/\Lambda\right)^{1/2}\sinh\left[\left(\Lambda/3
\right)^{1/2}t\right], \eqno(3.2b)$$
for $\Lambda>0$, and
$$R\left(t\right)=\left(-3/\Lambda\right)^{1/2}\cosh\left[\left(-\Lambda/3
\right)^{1/2}t\right] \eqno(3.2c)$$
for $\Lambda<0$. These functions are plotted in Fig. 3.1, curves (b) and (c),
respectively. It will be noted that both universes begin with a big bang at
$t=0$. The first of these curves expands monotonically, whereas the second one
oscillates. In our actual universe, the mass density near the singularity at
$t=0$ was extremely large, and hence this model cannot be used to describe
its behavior near this time.

Finally, for the universe with a flat three-dimensional space the Einstein 
field equations yield the analog to Eq. (3.1a),
$$3\left(dR/dt\right)^2/R^2=\Lambda.\eqno(3.1c)$$
This equation has meaning only for $\Lambda>0$, and it has the solution
$$R\left(t\right)=R\left(0\right)\exp\left[\left(\Lambda/3
\right)^{1/2}t\right]. \eqno(3.2d)$$
This universe expands exponentially. This model, described by Eq. (3.2d), is
usually called the {\it de Sitter universe}.
\section{Remarks of the Standard Theory}
To conclude the discussion on the Friedmann and Lema\^\i tre universes, we 
briefly discuss the case in which both the matter density and the cosmological
constant are not zero. It will be noted that exact solutions of the 
differential equations describing the expansion of the universe in that case,
given by Eqs. (2.8), are not known. These general models can be thought of as
a combination of the Friedmann and Lema\^\i tre models.

Consider a universe that begins with a big bang. At an early time, the 
universe must have been very dense, and we can neglect the cosmological term. 
Hence, we have approximately a Friedmann universe. As the universe expands
and the mass density decreases, the cosmological term will become more
important. In the Friedmann models of zero and negative curvature, the 
universe expands monotonically and the decrease in mass density is monotonic 
too. The cosmological term will ultimately dominate the behavior of the
universe, and it gradually turns into an empty Lema\^\i tre universe with zero
or negative curvature. In the case of negative curvature with $\Lambda<0$,
the expansion will stop at some later time, reverses and finally ends up in a
re-contracting universe of negative curvature.

In the case of a Friedmann universe with a positive curvature, the mass 
density reaches minimum when the radius of curvature is at its maximum. Hence,
the cosmological term will dominate the behavior of the universe only if it is
sufficiently large compared with the minimum mass density. The critical value
of $\Lambda$ is given by $\Lambda_E=(\pi/2GM)^2$. If $\Lambda$ is larger than
$\Lambda_E$, then the Friedmann universe with positive curvature gradually 
turns into an expanding Lema\^\i tre universe with positive curvature (Fig. 
4.1). In the case $\Lambda=\Lambda_E$, the transition is never completed and
the expansion stops at a constant value of $R=1/\Lambda_E^{1/2}$. This static
universe is called the Einstein universe. The universe at this value of $R$,
however, is not stable. Any perturbation in $R$ leads either to monotonic
expansion (toward an empty Lema\^\i tre model) or to contraction (toward a 
contracting Friedmann universe). 

In the final analysis, it follows that the expansion of the universe is
determined by the so-called cosmological parameters. These can be taken as the
mass density $\rho$, the Hubble constant $H$ and the deceleration parameter
$q$. In the following we give a brief review of these parameters and the 
relationship between them. In the rest of the paper we will concentrate on the
theory with dynamical variables that are actually measured by astronomers: 
distances, redshifts and the mass density.

Equations (2.8) can be written as
$$3\left(H^2+k/R^2\right)=8\pi G\rho+\Lambda,\eqno(4.1)$$
where $k=1$, $0$, or $-1$, for the cases of positive, zero, or negative
curvature, respectively. Using Eq. (4.1) and $\Omega=\rho/\rho_c$, where
$\rho_c=3H_0^2/8\pi G$, one obtains
$$\Omega=1+k/H^2R^2-\Lambda/3H^2.\eqno(4.2)$$
It follows from these equations that the curvature of the universe is
determined by $H$, $\rho$ and $\Lambda$, or equivalently, $H$, $\Omega$ and 
$\Lambda$:
$$\Omega>1-\Lambda/3H^2,\eqno(4.3a)$$
for positive curvature,
$$\Omega<1-\Lambda/3H^2,\eqno(4.3b)$$
for negative curvature, and
$$\Omega=1-\Lambda/3H^2,\eqno(4.3c)$$
for zero curvature.

The deceleration parameter is defined as
$$q\equiv -\left[1+\left(1/H^2\right)dH/dt\right],\eqno(4.4)$$
and it can be shown that
$$q=\Omega/2-\Lambda/3H^2.\eqno(4.5)$$
Using Eq. (4.5) we can eliminate $\Lambda$ from Eqs. (4.3) and obtain
$$3\Omega/2>1+q,\eqno(4.6a)$$
for positive curvature,
$$3\Omega/2<1+q,\eqno(4.6b)$$
for negative curvature, and
$$3\Omega/2=1+q,\eqno(4.6c)$$
for zero curvature.

It is worthwhile mentioning that one of the Friedmann theory assumptions is 
that the type of the universe is determined by $\Omega=
\rho/\rho_c$, where $\rho_c=3H_0^2/8\pi G$, which requires that the sign of 
$(\Omega-1)$ must not change throughout the evolution of the universe so as to 
change the kind of the universe from one to another. That means in this 
theory, the universe has only one kind of curvature throughout its evolution
and does not permit going from one curvature into another. In other words the 
universe has been and will be in only one
form of expansion. It is not obvious, 
however, that this is indeed a valid assumption whether theoretically or
experimentally. As will be shown in the sequel, the universe has actually
three phases of expansion, and it {\it does} go from one to the second and 
then to the third phase.

In the combined Friedmann-Lema\^\i tre theory discussed above in which both
the matter density and the cosmological constant are not zero, nevertheless,
the theory does permit the change of sign of the decelerating parameter $q$,
as can be seen from Fig. 4.1. There exist no equations, however, that describe 
this kind of transfer from one type of universe to another.
\section{Cosmological Theory in Terms of Distance and Redshift}
A new outlook at the universe expansion can be achieved and is presented here.
The new theory has the following features: (1) It gives a direct relationship
between distances and redshifts. (2) It is fully general relativistic. (3) It
includes two universal constants, the speed of light in vacuum $c$, and the 
Hubble time in the absence of gravity $\tau$ (might also be called the {\it
Hubble time in vacuum}). (4) The redshift parameter $z$ is taken as the 
time-like coordinate. (5) The energy-momentum tensor is represented 
differently by including in it a term which is equivalent to the cosmological
constant. And (6) it predicts that the universe has three phases of expansion: 
decelerating, constant and accelerating, but it is now in the stage of 
accelerating expansion phase after having gone through the other two phases. 

Our starting point is Hubble's cosmological equation of state, Eq. (1.1). One
can keep the velocity $v$ in equation (1.1) or replace it with the redshift 
parameter $z$ by means of $z=v/c$. Since {\bf R}$=(x_1,\mbox{ }x_2,\mbox{ }
x_3)$, the square of Eq. (1.1) then yields
$$c^2H_0^{-2}z^2-\left(x_1^2+x_2^2+x_3^2\right)=0.\eqno(5.1)$$
Our aim is to write our equations in an invariant way so as to enable us to
extend them to curved space. Equation (5.1) is not invariant since $H_0^{-1}$
is the Hubble time at present. At the limit of zero gravity, Eq. (5.1) will
have the form
$$c^2\tau^2z^2-\left(x_1^2+x_2^2+x_3^2\right)=0,\eqno(5.2)$$ 
where $\tau$ is Hubble's time in vacuum, which is a {\it universal constant}
the numerical value of which will be determined in the sequel by relating it
to $H_0^{-1}$ at different situations. Equation (5.2) provides the basis of a
cosmological special relativity and was investigated extensively [11-16].

In order to make Eq. (5.2) adaptable to curved space we write it in a 
differential form:
$$c^2\tau^2dz^2-\left(dx_1^2+dx_2^2+dx_3^2\right)=0,\eqno(5.3)$$ 
or, in a covariant form in flat space,
$$ds^2=\eta_{\mu\nu}dx^\mu dx^\nu =0,\eqno(5.4a)$$
where $\eta_{\mu\nu}$ is the ordinary Minkowskian metric, and our coordinates
are $(x^0,\mbox{ }x^1,\mbox{ }x^2,\mbox{ }x^3)=(c\tau z,\mbox{ }x_1,\mbox{ }
x_2,\mbox{ }x_3)$. Equation (5.4a) expresses the null condition, familiar from
light propagation in space, but here it expresses the universe expansion in
space. The generalization of Eq. (5.4a) to a covariant form in curved space 
can immediately be made by replacing the Minkowskian metric $\eta_{\mu\nu}$
by the curved Riemannian geometrical metric $g_{\mu\nu}$,
$$ds^2=g_{\mu\nu}dx^\mu dx^\nu =0,\eqno(5.4b)$$ 
obtained from solving the Einstein field equations.

Because of the spherical symmetry nature of the universe, the metric we seek 
is of the form [8]     
$$ds^2=c^2\tau^2dz^2-e^\lambda dr^2-r^2\left(d\theta^2+\sin^2\theta d\phi^2
\right),\eqno(5.5)$$
where co-moving coordinates, as in the Friedmann theory, are used and 
$\lambda$ is a function of the radial distance $r$. The metric (5.5) is static
and solves the Einstein field equation (2.7). When looking for static 
solutions, Eq. (2.7) can also be written as
$$e^{-\lambda}\left(\lambda '/r-1/r^2\right)+1/r^2=8\pi GT_0^0,\eqno(5.6)$$
when $\Lambda$ is taken zero, and where a prime denotes differentiation with 
respect to $r$. 

In general relativity theory one takes for $T_0^0=\rho$. So is the situation 
in Newtonian gravity where one has the Poisson equation $\nabla^2\phi=4\pi G
\rho$. At points where $\rho=0$ one solves the vacuum Einstein field equations
and the Laplace equation $\nabla^2\phi=0$ in Newtonian gravity. In both 
theories a null (zero) solution is allowed as a trivial case. In cosmology, 
however, there exists no situation at which $\rho$ can be zero because the
universe is filled with matter. In order to be able to have zero on the
right-hand side of Eq. (5.6) we take $T_0^0$ not as equal to $\rho$ but to
$\rho-\rho_c$, where $\rho_c$ is chosen by us now as a {\it constant} given
by $\rho_c=3/8\pi G\tau^2$. 

The introduction of $\rho_c$ in the energy-momentum tensor might be regarded 
as adding a cosmological constant to the Einstein field equations. But this 
is not exactly so, since the addition of $-\rho_c$ to $T_0^0$ means also 
fixing the numerical value of the cosmological constant and is no more a
variable to be determined by experiment. At any rate our reasons are 
philosophically different from the standard point of view, and this approach
has been presented and used in earlier work [17].

The solution of Eq. (5.6), with $T_0^0=\rho-\rho_c$, is given by
$$e^{-\lambda}=1-\left(\Omega-1\right)r^2/c^2\tau^2,\eqno(5.7)$$
where $\Omega=\rho/\rho_c$. Accordingly, if $\Omega>1$ we have $g_{rr}=-\left(
1-r^2/R^2\right)^{-1}$, where
$$R^2=c^2\tau^2/\left(\Omega-1\right),\eqno(5.8a)$$ 
exactly equals to $g_{11}$ given by Eq. (2.4) for the positive-curvature 
Friedmann universe that is obtained in the standard theory by purely 
geometrical manipulations (see Sect. 2). If $\Omega<1$, we can write $g_{rr}=
-\left(1+r^2/R^2\right)^{-1}$ with
$$R^2=c^2\tau^2/\left(1-\Omega\right),\eqno(5.8b)$$ 
which is equal to $g_{11}$ given by Eq. (2.5) for the negative-curvature 
Friedmann universe. In the above equations $r<R$.

In Fig. 5.1 a plot of $R$ as a function of $\Omega$, according to Eqs. (5.8), 
is given. One can interpret $R$ as the boundary of the universe within which
matter can exist, although it is not necessarily that the matter fills up all 
the 
space bounded by $R$.

Moreover, we know that the Einstein field equations for these cases are given
by Eqs. (2.8) which, in our new notation, have the form
$$\left[\left(dR/dz\right)^2+c^2\tau^2\right]/R^2=\left(\Omega-1\right),
\eqno(5.9a)$$
$$\left[\left(dR/dz\right)^2-c^2\tau^2\right]/R^2=\left(\Omega-1\right).
\eqno(5.9b)$$
As is seen from these equations, if one neglects the first term in the square
brackets with respect to the second ones, $R^2$ will be exactly reduced to
their values given by Eqs. (5.9).

The expansion of the universe can now be determined from the null condition 
$ds=0$, Eq. (5.4b), using the metric (5.5). Since the expansion is radial, one
has $d\theta=d\phi=0$, and the equation obtained is
$$dr/dz=c\tau\left[1+\left(1-\Omega\right)r^2/c^2\tau^2\right]^{1/2}.
\eqno(5.10)$$

The second term in the square bracket of Eq. (5.10) represents the deviation
from constant expansion due to gravity. For without this term, Eq. (5.10)
reduces to $dr/dz=c\tau$ or $dr/dv=\tau$, thus $r=\tau v+$constant. The 
constant can be taken zero if one assumes, as usual, that at $r=0$ the 
velocity should also vanish. Accordingly we have $r=\tau v$ or $v=\tau^{-1}r$.
When $\Omega=1$, namely when $\rho=\rho_c$, we have a constant expansion.

The equation of motion (5.10) can be integrated exactly by the substitutions
$$\begin{array}{cc}\sin\chi=\alpha r/c\tau;&\Omega>1,\\\end{array}\eqno(5.11a)$$
$$\begin{array}{cc}\sinh\chi=\beta r/c\tau;&\Omega<1,\\\end{array}\eqno(5.11b)$$
where 
$$\begin{array}{cc}\alpha=\left(\Omega-1\right)^{1/2},&
\beta=\left(1-\Omega\right)^{1/2}.\\\end{array}\eqno(5.12)$$

For the $\Omega>1$ case a straightforward calculation, using Eq. (5.11a), gives
$$dr=\left(c\tau/\alpha\right)\cos\chi d\chi,\eqno(5.13)$$
and the equation of the universe expansion (5.10) yields
$$d\chi=\alpha dz.\eqno(5.14a)$$
The integration of this equation gives
$$\chi=\alpha z+\mbox{\rm constant}.\eqno(5.15a)$$
The constant can be determined, using Eq. (5.11a). For at $\chi=0$, we have 
$r=0$ and $z=0$, thus
$$\chi=\alpha z,\eqno(5.16a)$$
or, in terms of the distance, using (5.11a) again,
$$\begin{array}{cc}r\left(z\right)=\left(c\tau/\alpha\right)\sin\alpha z;&
\alpha=\left(\Omega-1\right)^{1/2}.\\\end{array}\eqno(5.17a)$$
This is obviously a decelerating expansion.

For $\Omega<1$, using Eq. (5.11b), then a similar calculation yields for the
universe expansion (5.10)
$$d\chi=\beta dz,\eqno(5.14b)$$
thus
$$\chi=\beta z+\mbox{\rm constant}.\eqno(5.15b)$$ 
Using the same initial conditions used above then give
$$\chi=\beta z,\eqno(5.16b)$$  
and in terms of distances,
$$\begin{array}{cc}r\left(z\right)=\left(c\tau/\beta\right)\sinh\beta z;&
\beta=\left(1-\Omega\right)^{1/2}.\\\end{array}\eqno(5.17b)$$
This is now an accelerating expansion.

For $\Omega=1$ we have, from Eq. (5.10), 
$$d^2r/dz^2=0.\eqno(5.14c)$$
The solution is, of course,
$$r\left(z\right)=c\tau z.\eqno(5.17c)$$
This is a constant expansion.

It will be noted that the last solution can also be obtained directly from
the previous two ones for $\Omega>1$ and $\Omega<1$ by going to the limit
$z\rightarrow 0$, using L'Hospital lemma, showing that our solutions are
consistent. It will be shown later on that the constant expansion is just a
transition stage between the decelerating and the accelerating expansions as 
the universe evolves toward its present situation. 

Figure 5.2 describes the Hubble diagram of the above solutions for the three
types of expansion for values of $\Omega$ from 100 to 0.24. The figure 
describes the three-phase evolution of the universe. Curves (1) to (5) 
represent the stages of  decelerating expansion according to Eq. (5.17a). As
the density of matter $\rho$ decreases, the universe
goes over from the lower curves to the upper ones, and it does not have
enough time to close up to a big crunch. The universe subsequently goes to
curve (6) with $\Omega=1$, at which time it has a constant expansion for a
fraction of a second. This then followed by going to the upper curves (7) and
(8) with $\Omega<1$ where the universe expands with acceleration according to
Eq. (5.17b). A curve of this kind fits the present situation of the universe.
For curves (1) to (4) in the diagram we use the cut off when the curves were
at their maximum (or the same could be done by using the cut off as determined
by $R$ of Fig. 5.1). In Table 5.1 we present the cosmic times with respect to
the big bang and the cosmic radiation temperature for each of the curves 
appearing in Fig. 5.2.

In order to decide which of the three cases is the appropriate one at the
present time, we have to write the solutions (5.17) in the ordinary Hubble law
form $v=H_0r$. To this end we change variables from the redshift parameter $z$
to the velocity $v$ by means of $z=v/c$ for $v$ much smaller than $c$. For
higher velocities this relation is not accurate and one has to use a Lorentz 
transformation in order to relate $z$ to $v$. A simple calculation then shows
that, for receding objects, one has the relations
$$z=\left[\left(1+v/c\right)/\left(1-v/c\right)\right]^{1/2}-1,\eqno(5.18a)$$ 
$$v/c=z\left(z+2\right)/\left(z^2+2z+2\right).\eqno(5.18b)$$
We will assume that $v\ll c$ and consequently Eqs. (5.17) have the forms
$$r\left(v\right)=\left(c\tau/\alpha\right)\sin\left(\alpha v/c\right),
\eqno(5.19a)$$
$$r\left(v\right)=\left(c\tau/\beta\right)\sinh\left(\beta v/c\right),
\eqno(5.19b)$$
$$r\left(v\right)=\tau v.\eqno(5.19c)$$
Expanding now Eqs. (5.19a) and (5.19b) and keeping the appropriate terms, then 
yields
$$r=\tau v\left(1-\alpha^2v^2/6c^2\right),\eqno(5.20a)$$
for the $\Omega>1$ case, and
$$r=\tau v\left(1+\beta^2v^2/6c^2\right),\eqno(5.20b)$$  
for $\Omega<1$. Using now the expressions for $\alpha$ and $\beta$, given by 
Eq. (5.12), in Eqs. (5.20) then both of the latter can be reduced into a single 
equation
$$r=\tau v\left[1+\left(1-\Omega\right)v^2/6c^2\right].\eqno(5.21)$$
Inverting now this equation by writing it in the form $v=H_0r$, we obtain in 
the lowest approximation for $H_0$ the following:
$$H_0=h\left[1-\left(1-\Omega\right)v^2/6c^2\right],\eqno(5.22)$$
where $h=\tau^{-1}$. Using $v\approx r/\tau$, or $z=v/c$, we also obtain
$$H_0=h\left[1-\left(1-\Omega\right)r^2/6c^2\tau^2\right]
=h\left[1-\left(1-\Omega\right)z^2/6\right].\eqno(5.23)$$

Consequently, $H_0$ depends on the distance, or equivalently, on the redshift.
As is seen, $H_0$ has meaning only for $r\rightarrow 0$ or $z\rightarrow 0$,
namely when measured {\it locally}, in which case it becomes $h$.
\section{Concluding Remarks}
In recent years observers have argued for values of $H_0$ as low as 50 and as 
high as 90 km/sec-Mpc, some of the recent ones show $80\pm 17$ km/sec-Mpc 
[18-26]. There are the so-called ``short" and ``long" distance scales, with 
the higher and the lower values for $H_0$ respectively [27]. Indications are 
that the longer the distance of measurement the smaller the value of $H_0$.
By Eqs. (5.22) and (5.23) this is possible only for the case in which 
$\Omega<1$, namely when the universe is at an accelerating expansion.
Figures 5.3 and 5.4 show the Hubble diagrams for the predicted by theory 
distance-redshift relationship for the accelerating expanding universe at 
present time, whereas figures 5.5 and 5.6 show the experimental results [28-29].

Our estimate for $h$, based on published data, is $h\approx 85-90$ km/sec-Mpc.
Assuming $\tau^{-1}\approx 85$ km/sec-Mpc, Eq. (5.23) then gives
$$H_0=h\left[1-1.3\times 10^{-4}\left(1-\Omega\right)r^2\right],\eqno(6.1)$$
where $r$ is in Mpc. A computer best-fit can then fix both $h$ and $\Omega$.

To summarize, a new general-relativistic theory of cosmology has been 
presented in which the dynamical variables are those of Hubble's, i.e.
distances and redshifts. The theory describes the universe as having a 
three-phase evolution with a decelerating expansion, followed by a constant
and an accelerating expansion, and it predicts that the universe is now in the 
latter phase. As the density of matter decreases, while the universe is at the
decelerating phase, it does not have enough time to close up to a big crunch. 
Rather, it goes to the constant expansion phase, and then to the accelerating
stage.

The idea to express cosmological theory in terms of directly-measurable
quantities, such as distances and redshifts, was partially inspired by
Albert Einstein's favourite remarks on the theory of thermodynamics in his
Autobiographical Notes [30].

We wish to thank Professor Michael Gedalin for his help in preparing the
diagrams.

\newpage
\begin{center}{FIGURE CAPTIONS}\end{center}
Fig. 2.1 Radius of curvature of the positive-curvature Friedmann universe as
a function of time. The curve is a cycloid.\newline
Fig. 2.2 Radius of curvature of the negative-curvature Friedmann universe as
a function of time.\newline
Fig. 2.3 Radius of curvature of the flat Friedmann universe as
a function of time.\newline
Fig. 3.1 Radius of curvature of the empty Lema\^\i tre universes as a function of
time. (a) Positive-curvature model, $\Lambda >0$. (b) Negative-curvature model,
$\Lambda >0$. (c) Negative-curvature model, $\Lambda <0$. (d) Flat model, 
$\Lambda >0$.\newline
Fig. 4.1 Radius of curvature of a nonempty Lema\^\i tre universe, with 
$\Lambda >\Lambda_E$.\newline
Fig. 5.1 A plot of $R$ as a function of $\Omega$ according to equations (5.8).
$R$ is the boundary of the universe withinwhich matter can exist, although
it is not necessarily that the matter fills up all the space bounded by $R$.
\newline    
Fig. 5.2 Hubble's diagram describing the three-phase  evolution of the
universe according to Einstein's general relativity theory. Curves (1) to (5)
represent the stages of {\it decelerating} expansion according to $r(z)=
(c\tau/\alpha)\sin\alpha z$, where $\alpha=(\Omega -1)^{1/2}$, $\Omega=\rho/
\rho_c$, with $\rho_c$ a {\it constant}, $\rho_c=3/8\pi G\tau^2$, and $c$ and
$\tau$ are the speed of light and the Hubble time in vacuum (both universal 
constants). As the density of matter $\rho$ decreases, the universe goes over 
from the lower curves to the upper ones, but it does not have enough time to
close up to a big crunch. The universe subsequently goes to curve (6) with
$\Omega=1$, at which time it has a {\it constant} expansion for a fraction of
a second. This then followed by going to the upper curves (7)-(8) with $\Omega
<1$ where the universe expands with acceleration according to $r(z)=(c\tau/
\beta)\sinh\beta z$, where $\beta=(1-\Omega)^{1/2}$. One of these last
curves fits the present situation of the universe.\newline
Fig. 5.3 Hubble's diagram of the universe at the present phase of evolution
with accelerating expansion.\newline
Fig. 5.4 Hubble's diagram describing decelerating, constant and accelerating
expansions in a logarithmic scale.\newline
Fig. 5.5 Distance vs. redshift diagram showing the deviation from a constant
toward an accelerating expansion. [Sourse: A. Riess {\it et al., Astron. J.}
{\bf 116}, 1009 (1998)].\newline
Fig. 5.6 Relative intensity of light and relative distance vs. redshift.
[Sourse: A. Riess {\it et al., Astron. J.}
{\bf 116}, 1009 (1998)].
\newpage
\begin{tabular}{ccccc}
\hline\hline
Curve No.&$\Omega$&Time in units of $\tau$&Time (sec)&Temperature (K)\\
\hline
1&100&$3.1\times 10^{-6}$&$1.1\times 10^{12}$&1114.0\\
2&25&$9.8\times 10^{-5}$&$3.6\times 10^{13}$&279.0\\
3&10&$3.0\times 10^{-4}$&$1.1\times 10^{14}$&111.0\\
4&5&$1.2\times 10^{-3}$&$4.4\times 10^{14}$&56.0\\
5&1.5&$1.3\times 10^{-2}$&$4.7\times 10^{15}$&17.0\\
6&1&$3.0\times 10^{-2}$&$1.1\times10^{16}$&11.0\\
7&0.5&$1.3\times 10^{-1}$&$4.7\times 10^{16}$&6.0\\
8&0.245&1.0&$3.6\times 10^{17}$&2.7\\
\hline\end{tabular}
\vspace{5mm}\newline
Table 5.1 {\small The cosmic times with respect to the big bang and the cosmic 
temperature for each of the curves appearing in Fig. 5.2. The calculations are
made using a Lorentz-like transformation that relates physical quantities at
different cosmic times when gravity is extremely weak [13].}

\newpage
\begin{figure}
\centering
\includegraphics{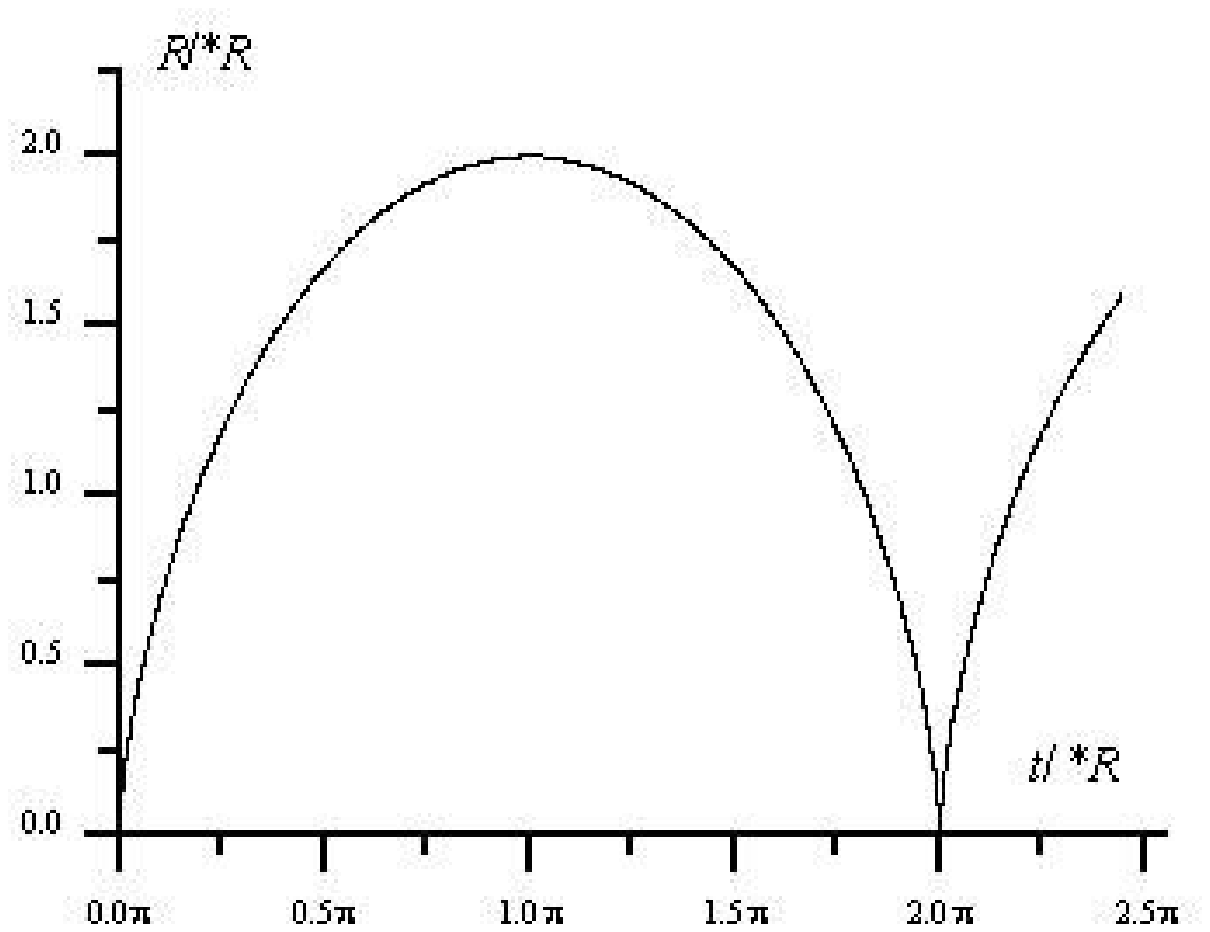}
\caption{}
\end{figure}

\newpage
\begin{figure}
\centering
\includegraphics{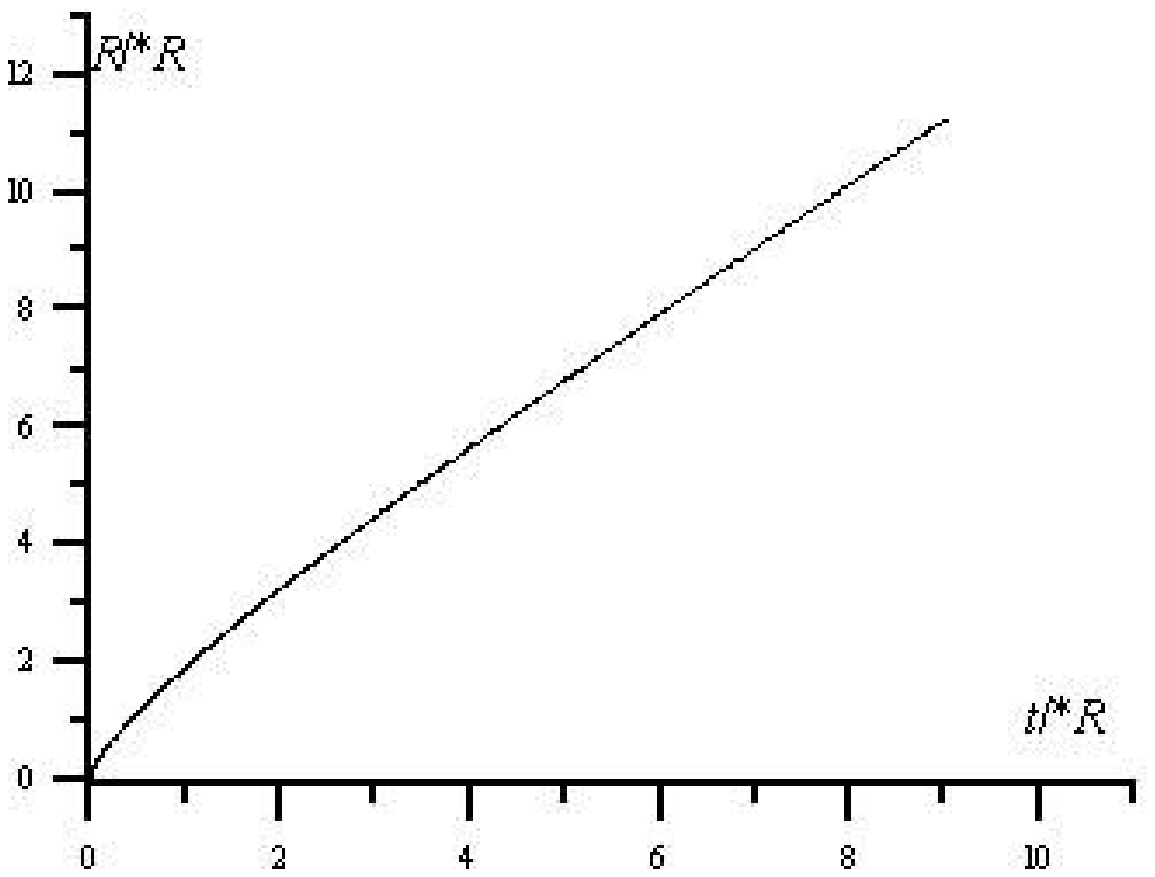}
\caption{}
\end{figure}

\newpage
\begin{figure}
\centering
\includegraphics{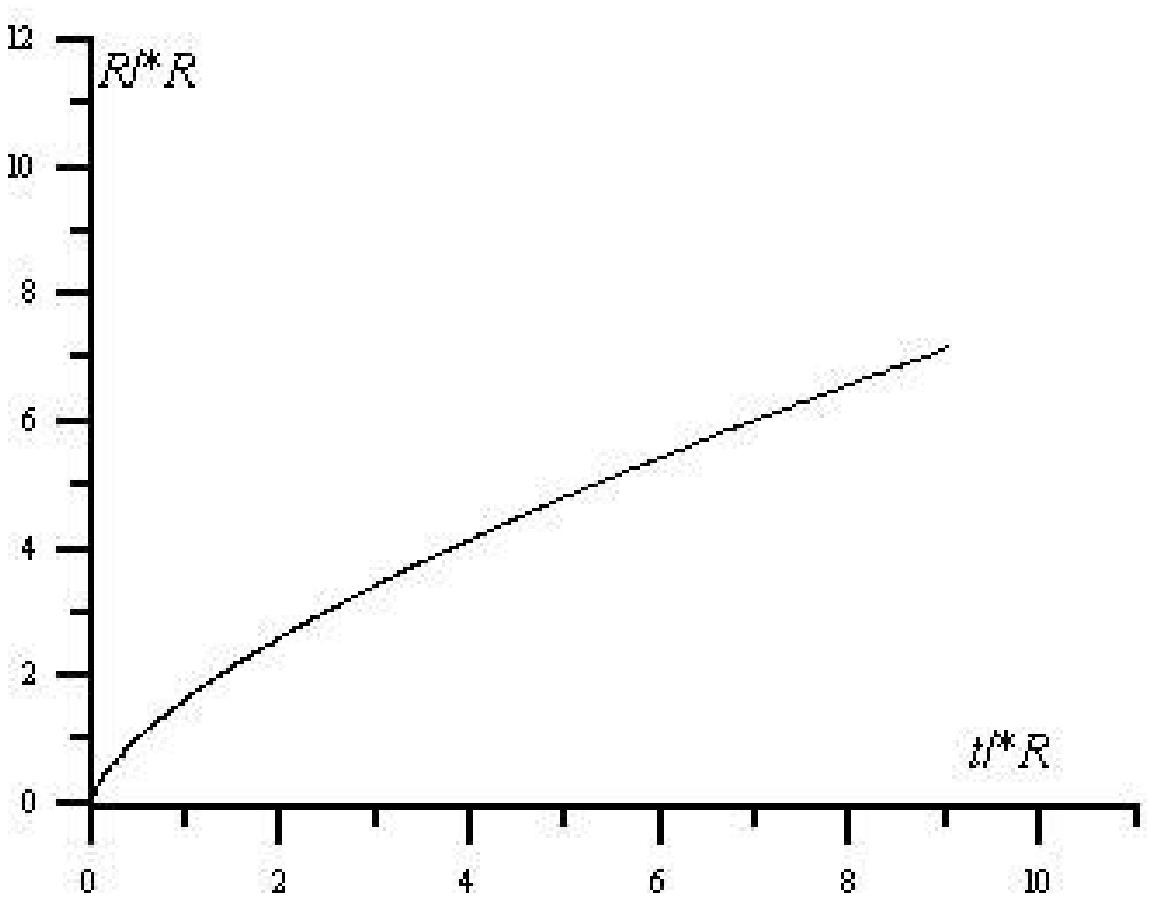}
\caption{}
\end{figure}

\addtocounter{newpart}{1}
\addtocounter{figure}{-3}
\newpage
\begin{figure}
\centering
\includegraphics{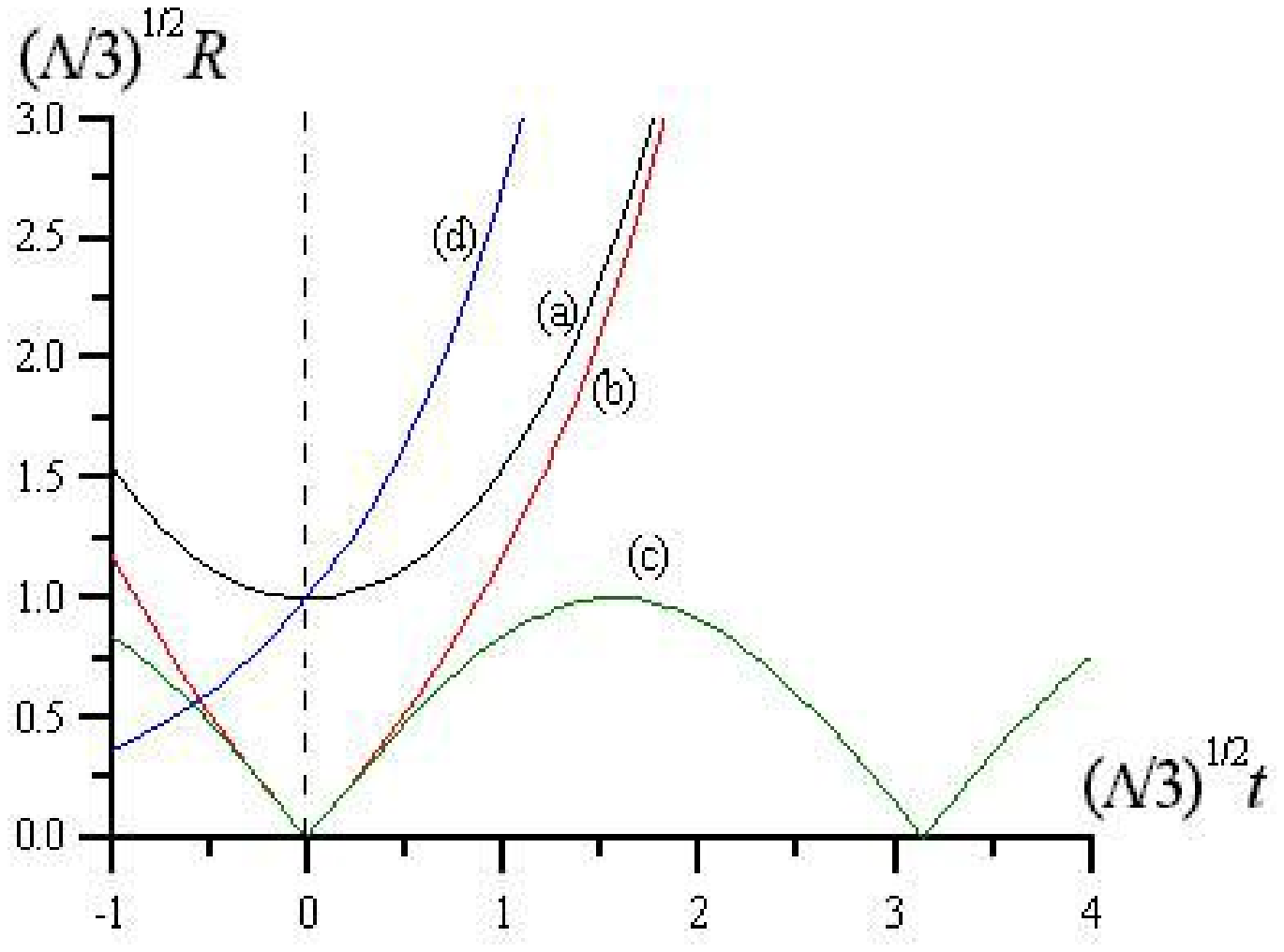}
\caption{}
\end{figure}

\addtocounter{newpart}{1}
\addtocounter{figure}{-1}
\newpage
\begin{figure}
\centering
\includegraphics{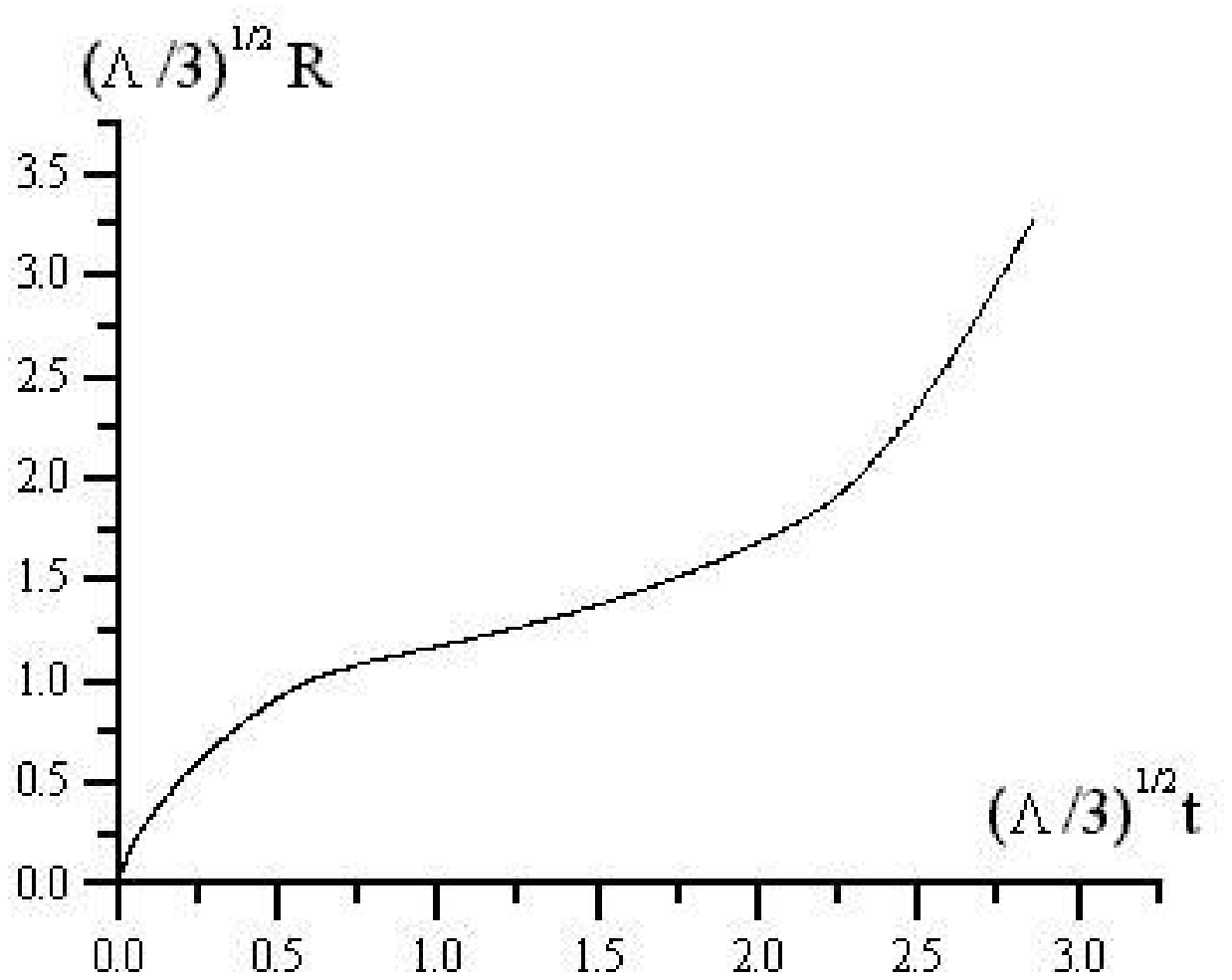}
\caption{}
\end{figure}

\addtocounter{newpart}{1}
\addtocounter{figure}{-1}
\newpage
\begin{figure}
\centering
\includegraphics{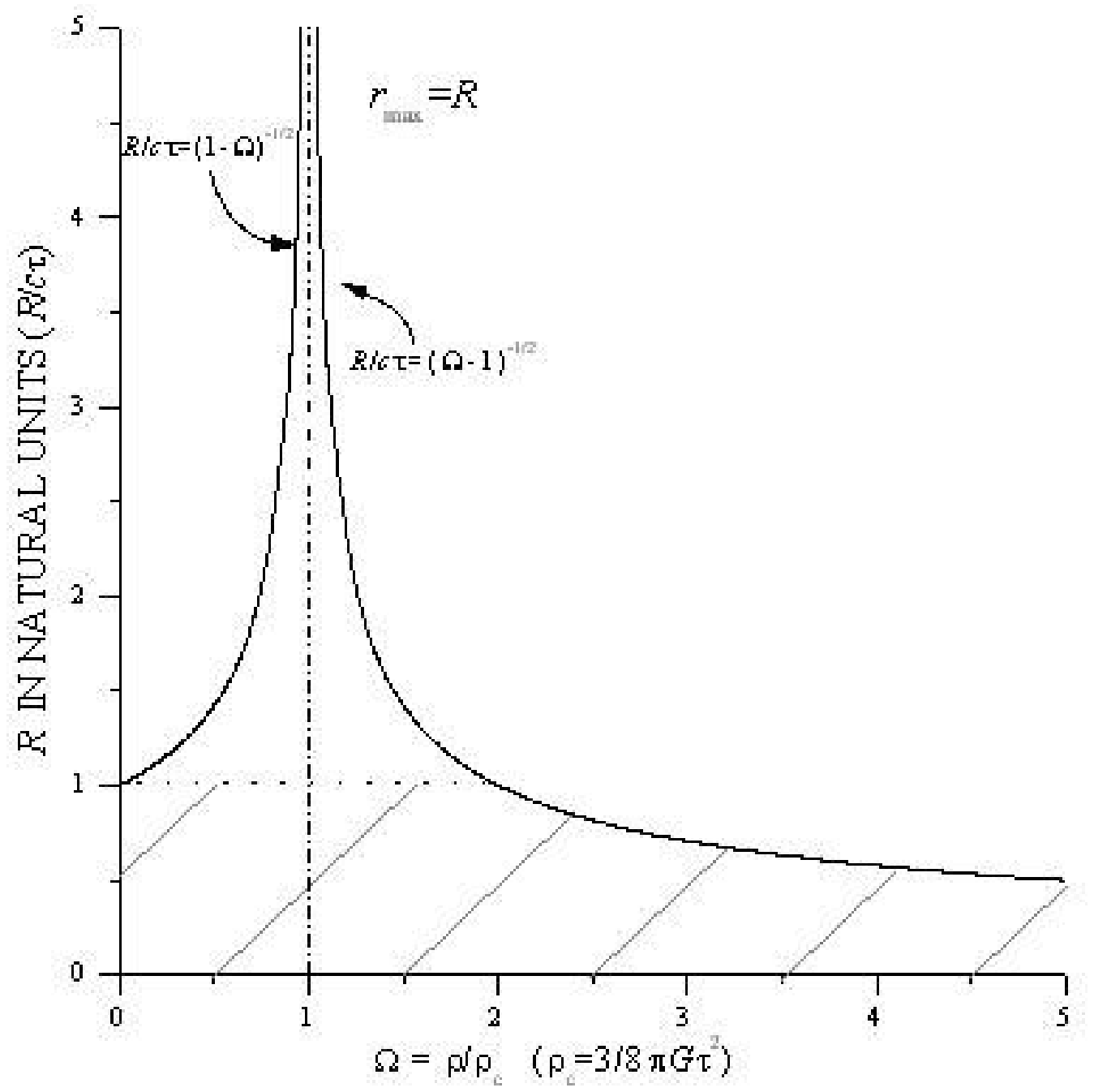}
\caption{}
\end{figure}

\newpage
\begin{figure}
\centering
\includegraphics{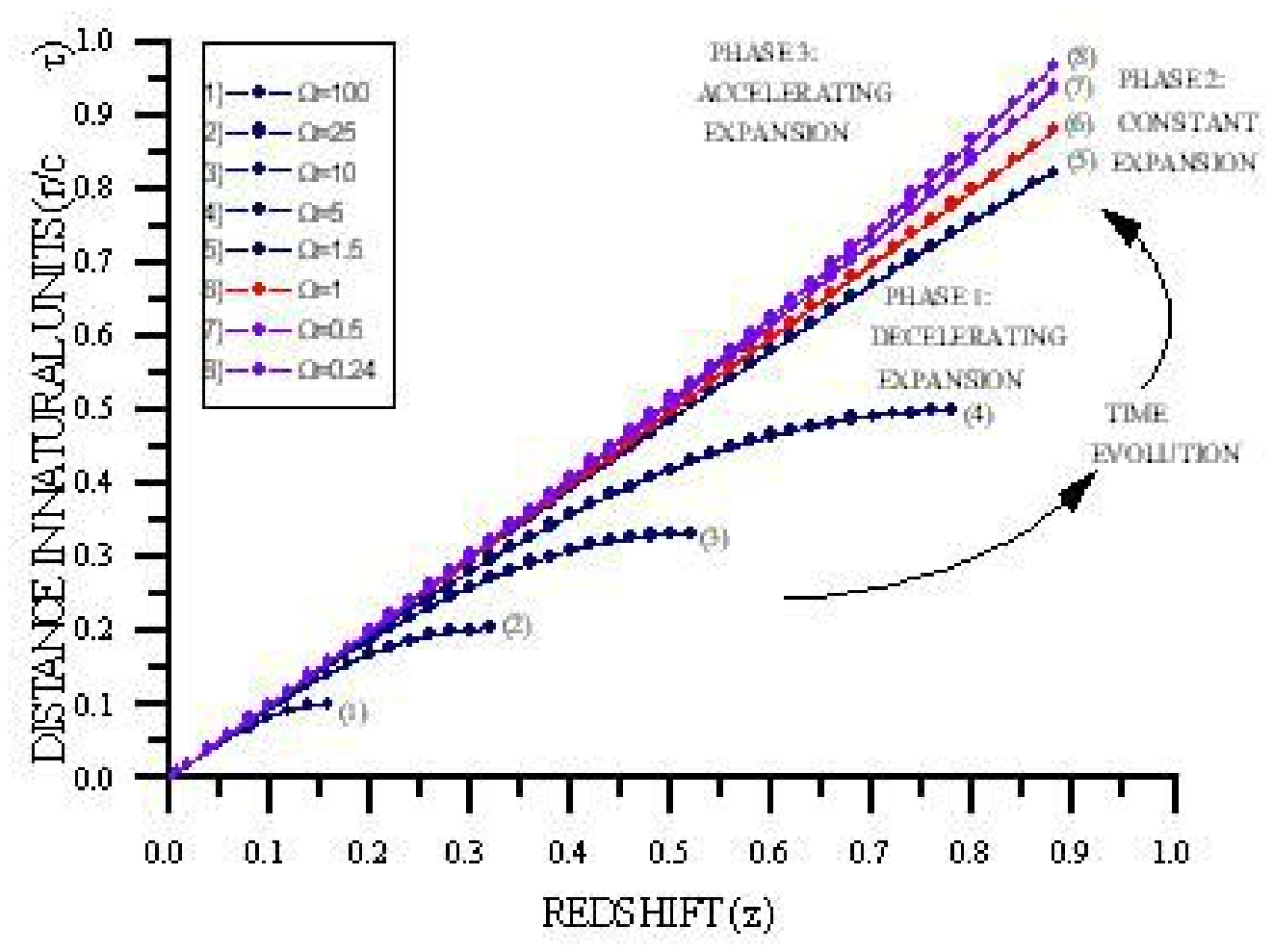}
\caption{}
\end{figure}

\newpage
\begin{figure}
\centering
\includegraphics{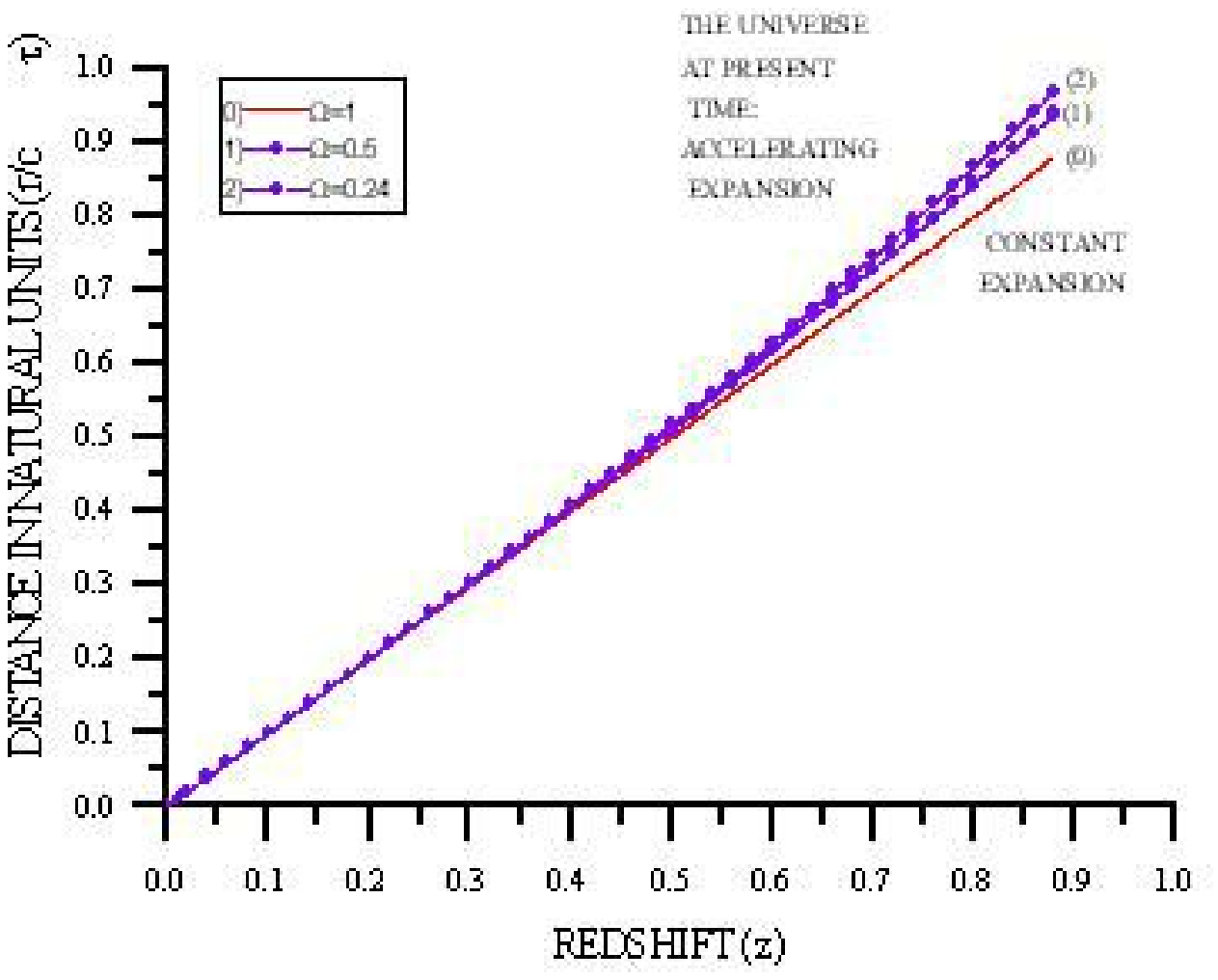}
\caption{}
\end{figure}

\newpage
\begin{figure}
\centering
\includegraphics{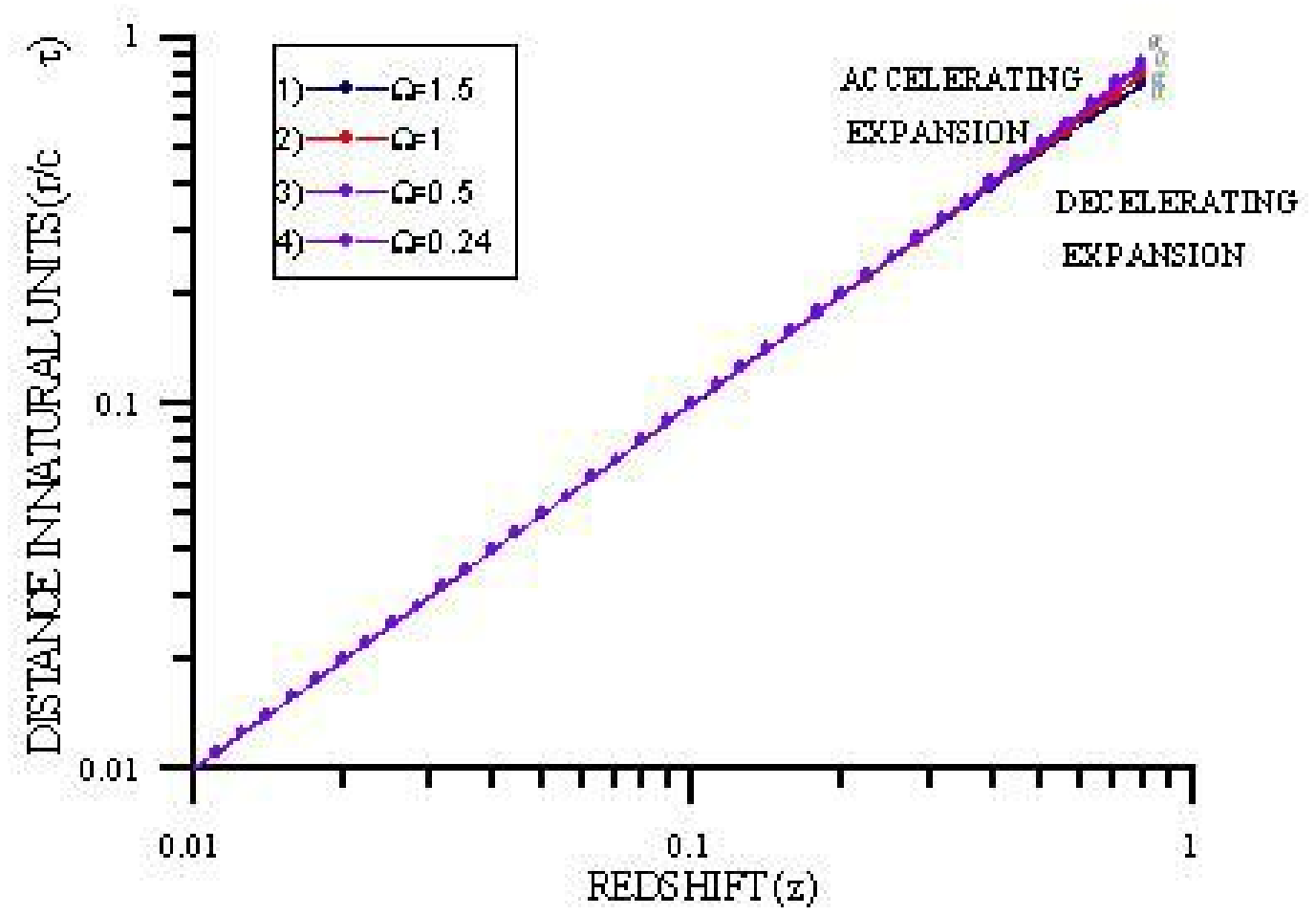}
\caption{}
\end{figure}

\newpage
\begin{figure}
\centering
\includegraphics{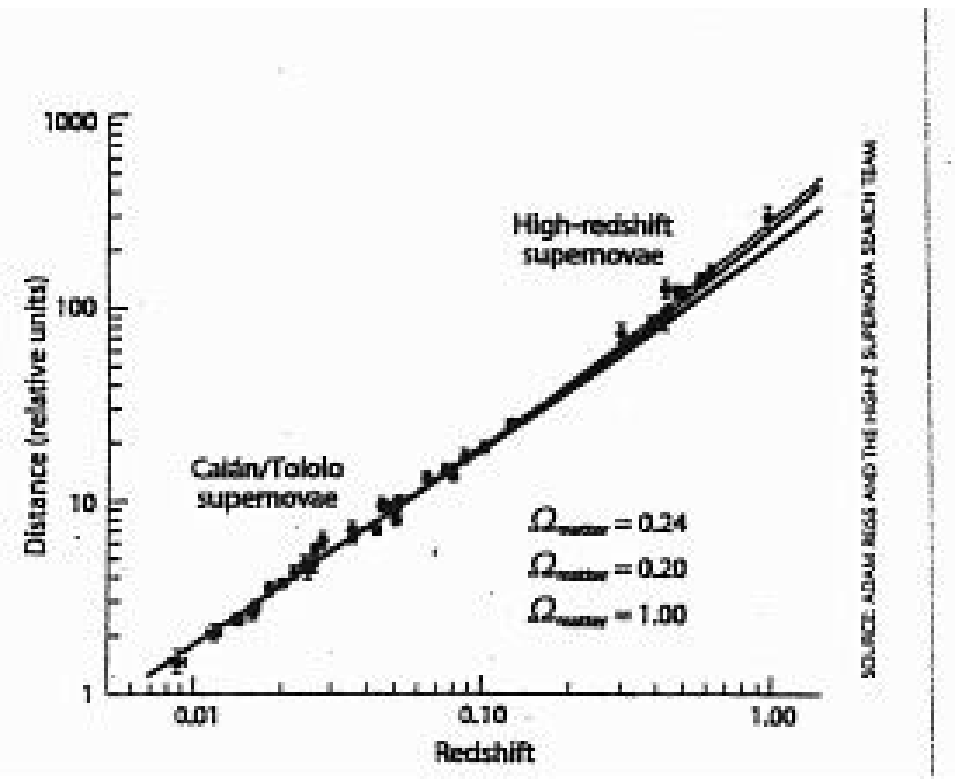}
\caption{}
\end{figure}

\newpage
\begin{figure}
\centering
\includegraphics{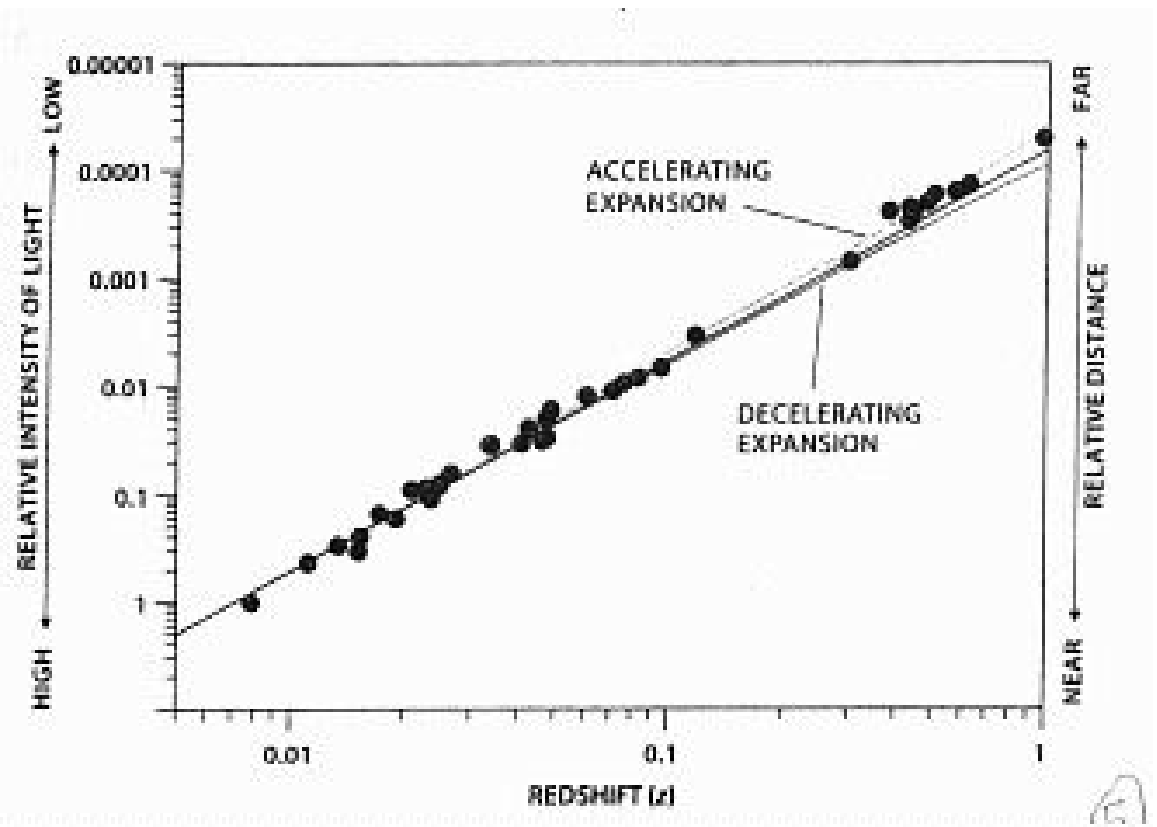}
\caption{}
\end{figure}

\end{document}